\documentclass[aps,prl,twocolumn,superscriptaddress]{revtex4-2}
\usepackage{amsmath,amssymb,amsbsy,graphicx,xcolor,times,subfigure,marvosym,color,bm,multirow,epstopdf}
\usepackage[latin1]{inputenc}
\begin{document}

\title{Gate-controlled anyon generation and detection in Kitaev spin liquids}

\author{G\'abor B. Hal\'asz}
\thanks{This manuscript has been authored by UT-Battelle, LLC, under contract DE-AC05-00OR22725 with the US Department of Energy (DOE). The publisher acknowledges the US government license to provide public access under the DOE Public Access Plan (http://energy.gov/downloads/doe-public-access-plan).}
\affiliation{Materials Science and Technology Division, Oak Ridge National Laboratory, Oak Ridge, TN 37831, USA}
\affiliation{Quantum Science Center, Oak Ridge, TN 37831, USA}


\begin{abstract}

Reliable manipulation of non-Abelian Ising anyons supported by Kitaev spin liquids may enable intrinsically fault-tolerant quantum computation. Here, we introduce a standalone scheme for both generating and detecting individual Ising anyons using tunable gate voltages in a heterostructure containing a non-Abelian Kitaev spin liquid and a monolayer semiconductor. The key ingredients of our setup are a Kondo coupling to stabilize an Ising anyon in the spin liquid around each electron in the semiconductor, and a large charging energy to allow control over the electron numbers in distinct gate-defined regions of the semiconductor. In particular, a single Ising anyon can be generated at a disk-shaped region by gate tuning its electron number to one, while it can be interferometrically detected by measuring the electrical conductance of a ring-shaped region around it whose electron number is allowed to fluctuate between zero and one. We provide concrete experimental guidelines for implementing our proposal in promising candidate materials like $\alpha$-RuCl$_3$.

\end{abstract}


\maketitle


\emph{Introduction.}---The Kitaev model on the honeycomb lattice provides an exactly solvable realization of non-Abelian topological order in an electrically insulating setting~\cite{Kitaev-2006}. While the ground state of the original Kitaev model is a gapless quantum spin liquid, the inclusion of a small magnetic field induces a fully gapped topological phase, the non-Abelian Kitaev spin liquid, that harbors Majorana fermions and Ising anyons. Due to their non-Abelian nature, the Ising anyons support nonlocal degrees of freedom that act like intrinsic quantum memories in the sense that their quantum states are protected from local perturbations up to exponentially long time scales. Moreover, these nonlocal quantum states can be manipulated in an inherently fault-tolerant manner by exchanging the Ising anyons or moving them around each other, which is the essence of topological quantum computation~\cite{Kitaev-2003, Nayak-2008}.

\begin{figure}[b]
\includegraphics[width=0.92\columnwidth]{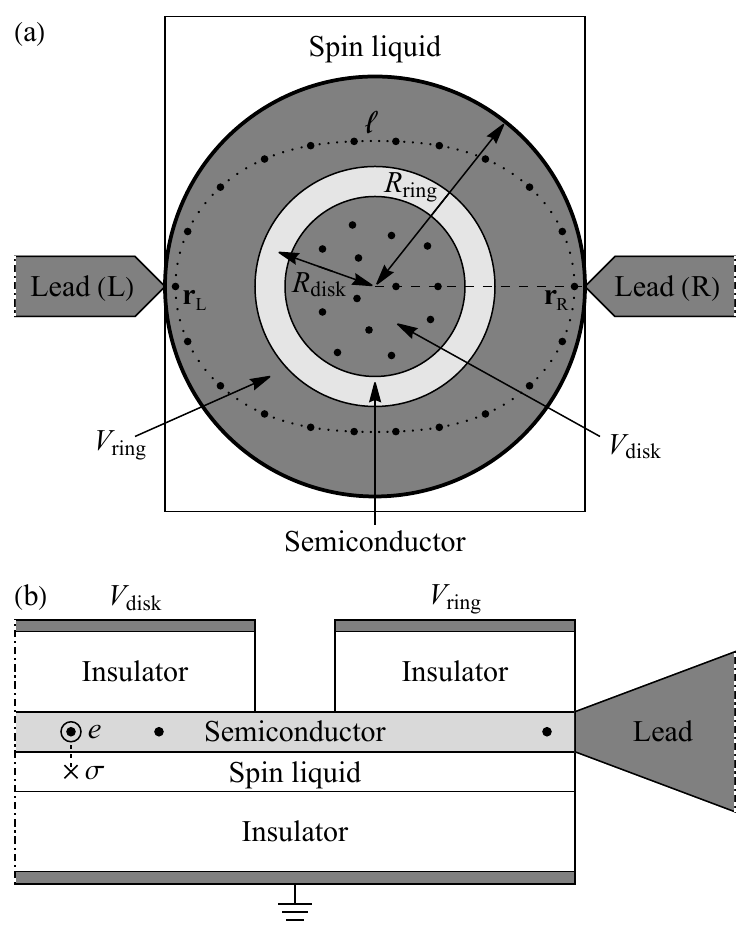}
\caption{Top view (a) and side view (b) of the proposed experimental setup with the cross section shown in (b) corresponding to the dashed line in (a). The monolayer semiconductor (light gray), whose boundary is denoted by a thick line in (a), is gated with two electrodes of voltages $V_{\mathrm{disk}}$ and $V_{\mathrm{ring}}$ above (dark gray). The semiconductor contains dopant sites (large dots) that (a) form a closed loop of length $\ell$ (dotted line) in the ring region with two sites $\mathbf{r}_{\Lambda}$ being in direct contact with the leads $\Lambda = \mathrm{L}, \mathrm{R}$, and (b) are able to localize electrons $e$ and, hence, Ising anyons $\sigma$ in the disk region.} \label{fig-1}
\end{figure}

Remarkably, in candidate materials of the non-Abelian Kitaev spin liquid, like $\alpha$-RuCl$_3$~\cite{Plumb-2014, Sandilands-2015, Sears-2015, Majumder-2015, Johnson-2015, Sandilands-2016, Banerjee-2016, Banerjee-2017, Do-2017} under an in-plane magnetic field~\cite{Kubota-2015, Leahy-2017, Sears-2017, Wolter-2017, Baek-2017, Banerjee-2018, Hentrich-2018, Jansa-2018, Kasahara-2018, Widmann-2019, Balz-2019, Yamashita-2020, Czajka-2021, Yokoi-2021, Bruin-2022, Czajka-2023}, the topological gap that protects quantum information could be as large as $10$ K, potentially facilitating quantum computation at elevated temperatures compared to other platforms. As an important first step toward anyon manipulation, it has also been recently understood that individual anyons in the Kitaev spin liquid can be stabilized around local defects like spin vacancies~\cite{Willans-2010, Willans-2011, Kao-2021} or dynamically generated in a magnetic tunnel junction~\cite{Liu-2022} and detected via anyonic edge interferometry~\cite{Aasen-2020, Klocke-2021, Wei-2021, Klocke-2022, Wei-2023} or scanning tunneling measurements~\cite{Feldmeier-2020, Pereira-2020, Konig-2020, Udagawa-2021, Bauer-2023, Takahashi-2022, Kao-2023}. In a real experiment, however, one cannot focus on anyon generation or detection only as reliably demonstrating either of them requires the other one, and it is not clear how the individual setups for anyon generation and detection above can be interfaced with each other.

In this Letter, we propose a standalone scheme for both generating and detecting non-Abelian Ising anyons using a gated semiconducting monolayer that is Kondo coupled to a non-Abelian Kitaev spin liquid. The main idea is that Ising anyons in the spin liquid can be bound to electrons in the neighboring semiconductor by the Kondo coupling~\cite{Dhockak-2010, Das-2016, Vojta-2016} while the electron numbers in various gated regions of the semiconductor can be accurately controlled by gate voltages in the limit of a large charging energy. In turn, each anyon stabilized at a disk-shaped region can be detected by measuring the electrical conductance of a ring-shaped region surrounding it, which corresponds to a bulk version of anyonic interferometry. Since our setup relies on tunable gate voltages, it is a natural generalization of analogous schemes in the context of topological superconductors~\cite{Alicea-2011, Aasen-2016, Karzig-2017, Lutchyn-2018}. As such, it offers a promising pathway toward scalable qubit architectures based on non-Abelian anyons of quantum spin liquids.

\emph{General setup.}---We consider the heterostructure in Fig.~\ref{fig-1} containing a single layer of a non-Abelian Kitaev spin liquid and a monolayer semiconductor that is only present in a circular region of radius $R_{\mathrm{ring}}$. The gating electrodes above the semiconductor define two sub-regions within this circular region, a disk inside radius $R_{\mathrm{disk}}$ and a ring between radii $R_{\mathrm{disk}}$ and $R_{\mathrm{ring}}$, that have individual gate voltages and charging energies, while the insulating layers suppress electron tunneling between the semiconductor, the gating electrodes above, and the grounded electrode below.

Since the semiconductor is connected to two metallic leads along the outer radius of the ring [see Fig.~\ref{fig-1}(a)], the number of electrons within each of the disk and ring regions can be controlled by the corresponding gate voltage. In turn, assuming a sufficiently strong Kondo coupling between the semiconductor and the spin liquid (to be specified later), each electron $e$ in the semiconducting layer binds a non-Abelian Ising anyon $\sigma$ in the neighboring spin-liquid layer [see Fig.~\ref{fig-1}(b)]. To stabilize these composite particles consisting of an electron and an Ising anyon, we also consider an array of electron dopants in the semiconductor that induce localized states for the electrons, thus slowing down their dynamics. Moreover, since the Ising anyons are gapped in the bulk and can only be created in pairs, it is important that the metallic leads are close to the edge of the spin liquid. This way, if an electron tunnels from the lead into the semiconductor, one Ising anyon can readily bind to the electron while the other one can be left behind at the spin-liquid edge with very little energy cost.

The Hamiltonian reads $H = H_{\mathrm{structure}} + H_{\mathrm{leads}} + H_{\mathrm{tunnel}}$, where $H_{\mathrm{structure}}$ governs the heterostructure itself,
\begin{equation}
H_{\mathrm{leads}} = \sum_{\Lambda = \mathrm{L}, \mathrm{R}} \sum_{k} \sum_{\mu = \uparrow, \downarrow} \zeta_k^{\phantom{\dag}} \hat{c}_{\Lambda, k, \mu}^{\dag} \hat{c}_{\Lambda, k, \mu}^{\phantom{\dag}} \label{eq-H-leads}
\end{equation}
describes free electrons in the left ($\mathrm{L}$) and right ($\mathrm{R}$) leads with momentum and spin indices $k$ and $\mu$, while
\begin{equation}
H_{\mathrm{tunnel}} = -\sum_{\Lambda = \mathrm{L}, \mathrm{R}} \, \sum_{\mathbf{r}, k, \mu} \left( T_{\Lambda, \mathbf{r}}^{\phantom{\dag}} c_{\mathbf{r}, \mu}^{\dag} \hat{c}_{\Lambda, k, \mu}^{\phantom{\dag}} + \mathrm{H.c.} \right) \label{eq-H-tunnel}
\end{equation}
corresponds to electron tunneling between the leads $\Lambda = \mathrm{L}, \mathrm{R}$ and the sites $\mathbf{r}$ of the semiconductor. In turn, the heterostructure Hamiltonian is $H_{\mathrm{structure}} = H_{\mathrm{semi}} + H_{\mathrm{Kitaev}} + H_{\mathrm{Kondo}}$, where $H_{\mathrm{semi}}$, $H_{\mathrm{Kitaev}}$, and $H_{\mathrm{Kondo}}$ describe electron dynamics in the semiconductor, spin dynamics in the spin liquid, and Kondo coupling between them, respectively. The Hamiltonian of the semiconducting layer,
\begin{align}
H_{\mathrm{semi}} = &-t \sum_{\langle \mathbf{r}, \mathbf{r}' \rangle} \sum_{\mu = \uparrow, \downarrow} \left( e^{i A_{\mathbf{r}, \mathbf{r}'}} c_{\mathbf{r}, \mu}^{\dag} c_{\mathbf{r}', \mu}^{\phantom{\dag}} + \mathrm{H.c.} \right) \nonumber \\
&- W \sum_{\mathbf{r} \in \mathbb{D}} \sum_{\mu = \uparrow, \downarrow} c_{\mathbf{r}, \mu}^{\dag} c_{\mathbf{r}, \mu}^{\phantom{\dag}} - \vec{h} \cdot \sum_{\mathbf{r}, \mu, \nu} c_{\mathbf{r}, \mu}^{\dag} \vec{\tau}_{\mu \nu}^{\phantom{\dag}} c_{\mathbf{r}, \nu}^{\phantom{\dag}} \nonumber \\
&+ \sum_{\eta = \mathrm{disk}, \mathrm{ring}} \left( E_{\eta} N_{\eta}^2 - e V_{\eta} N_{\eta} \right), \label{eq-H-semi}
\end{align}
contains an electron hopping amplitude $t$ with a Peierls phase factor $e^{i A_{\mathbf{r}, \mathbf{r}'}}$ representing an out-of-plane magnetic field, a binding energy $W$ at the dopant sites $\mathbf{r} \in \mathbb{D}$ [see Fig.~\ref{fig-1}], a Zeeman field $\vec{h}$ coupling to the total electron spin with the Pauli matrices $\vec{\tau} = (\tau^x, \tau^y, \tau^z)$, as well as charging energies $E_{\eta}$ and gate voltages $V_{\eta}$ that couple to the total electron numbers, $N_{\eta} = \sum_{\mathbf{r} \in \eta} \sum_{\mu} c_{\mathbf{r}, \mu}^{\dag} c_{\mathbf{r}, \mu}^{\phantom{\dag}}$, inside the disk and ring regions. The spin-liquid layer is governed by the exactly solvable Kitaev honeycomb Hamiltonian~\cite{Kitaev-2006}
\begin{equation}
H_{\mathrm{Kitaev}} = -K \sum_{\langle \mathbf{r}, \mathbf{r}' \rangle_{\alpha}} \sigma_{\mathbf{r}}^{\alpha} \sigma_{\mathbf{r}'}^{\alpha} - \kappa K \sum_{\langle \mathbf{r}, \mathbf{r}', \mathbf{r}'' \rangle_{\alpha \beta}} \sigma_{\mathbf{r}}^{\alpha} \sigma_{\mathbf{r}'}^{\gamma} \sigma_{\mathbf{r}''}^{\beta}, \label{eq-H-Kitaev}
\end{equation}
where $\sigma_{\mathbf{r}}^{\alpha}$ with $\alpha = x,y,z$ are spin-1/2 operators, $(\alpha \beta \gamma)$ is a general permutation of $(xyz)$, while $\langle \mathbf{r}, \mathbf{r}', \mathbf{r}'' \rangle_{\alpha \beta}$ is the path consisting of the two bonds $\langle \mathbf{r}, \mathbf{r}' \rangle_{\alpha}$ and $\langle \mathbf{r}', \mathbf{r}'' \rangle_{\beta}$ on the honeycomb lattice. The first term $\propto K$ corresponds to the original Kitaev model, in which gapless Majorana fermions are coupled to $\mathbb{Z}_2$ gauge fields, and the associated $\mathbb{Z}_2$ gauge fluxes are gapped excitations localized at hexagonal plaquettes. The second term $\propto \kappa K$ represents a magnetic field breaking time-reversal symmetry, which gaps out the Majorana fermions and turns the gauge fluxes into non-Abelian Ising anyons~\cite{Kitaev-2006}. Finally, the Kondo Hamiltonian
\begin{equation}
H_{\mathrm{Kondo}} = J \sum_{\mathbf{r}, \mu, \nu} \vec{\sigma}_{\mathbf{r}}^{\phantom{\dag}} \cdot \left( c_{\mathbf{r}, \mu}^{\dag} \vec{\tau}_{\mu \nu}^{\phantom{\dag}} c_{\mathbf{r}, \nu}^{\phantom{\dag}} \right) \label{eq-H-Kondo}
\end{equation}
introduces antiferromagnetic coupling between the localized spins in the spin liquid and the electron spins in the semiconductor. Here we assume for simplicity that the spin liquid and the semiconductor have the same honeycomb lattice and that their respective sites are directly on top of each other.

\emph{Anyon generation.}---In the rest of this work, we assume that the gate voltages $V_{\mathrm{disk}}$ and $V_{\mathrm{ring}}$ are tuned such that the disk region is deep inside a Coulomb-blockade valley with its electron number $N_{\mathrm{disk}}$ fixed to a small integer while the ring region is at a Coulomb-blockade peak with its electron number $N_{\mathrm{ring}}$ readily fluctuating between $0$ and $1$. Thus, if the disk is tuned into a neighboring Coulomb-blockade valley by changing $V_{\mathrm{disk}}$, an electron can move between the leads and the disk through the ring to facilitate the shift in $N_{\mathrm{disk}}$.

In the limit of infinitely large $W$, the low-energy eigenstates of $H_{\mathrm{structure}}$ have electrons localized at specific dopant sites $\mathbf{r} \in \mathbb{D}$ in the semiconductor, and the spins of these localized electrons then behave like Kondo impurities from the perspective of the spin liquid. For the gapless Kitaev spin liquid, corresponding to $\kappa = 0$ in Eq.~(\ref{eq-H-Kitaev}), it is well known that a single Kondo impurity induces a topological transition at a critical coupling $J \approx 0.35 K$ beyond which a gauge flux is attached to the Kondo impurity~\cite{Dhockak-2010, Das-2016, Vojta-2016}. In the Supplemental Material (SM)~\cite{SM}, we demonstrate that the same topological transition is also induced for the non-Abelian Kitaev spin liquid at a comparable critical coupling for any $0.1 \leq \kappa \leq 0.2$ even in the presence of a small Zeeman field $\vec{h}$. Furthermore, since the non-Abelian Kitaev spin liquid is gapped, the same conclusion is valid for multiple Kondo impurities as long as they are further apart than the correlation length.

Therefore, in the limit of $W \to \infty$, each electron localized in the semiconducting layer has a gauge flux bound to it in the neighboring spin-liquid layer, which in turn corresponds to a non-Abelian Ising anyon $\sigma$. We can then write a general low-energy eigenstate of $H_{\mathrm{structure}}$ for $N_{\mathrm{disk}} = N$ and $N_{\mathrm{ring}} = 0$ as $| N, \mathbf{R}, \chi \rangle$, where $\mathbf{R} = \{ \mathbf{r}_1, \ldots, \mathbf{r}_N \}$ specifies the positions of the electrons in the disk, while $\chi$ accounts for the nonlocal low-energy degrees of freedom that are spanned by the Ising anyons due to their non-Abelian nature~\cite{Kitaev-2006}.

For a finite but sufficiently large $W \gg t$, the electrons can tunnel between neighboring dopant sites $\mathbf{r} \in \mathbb{D}$ and $\mathbf{r}' \in \mathbb{D}$ through the non-dopant sites in between. If the tunneling process takes $s$ steps, the tunneling amplitude is on the order of $\tilde{t} \sim t^s / W^{s-1}$. Importantly, for a sufficiently small $\tilde{t} \ll K$, this electron tunneling process cannot create any bulk excitations in the spin liquid due to an excitation gap on the order of $K$. Therefore, the Ising anyons in the spin liquid must remain bound to their respective host electrons in the semiconductor even as the electrons move around the disk~\cite{Halasz-2014}.

In other words, the effective electron hopping of amplitude $\tilde{t}$ simply connects different low-energy states $| N, \mathbf{R}, \chi \rangle$. Thus, the ground state of $H_{\mathrm{structure}}$ for $N_{\mathrm{disk}} = N$ and $N_{\mathrm{ring}} = 0$ must take the general form
\begin{equation}
| \Omega_N \rangle = \sum_{\mathbf{R}, \chi} \omega_{N, \mathbf{R}, \chi} \, | N, \mathbf{R}, \chi \rangle, \label{eq-Omega}
\end{equation}
where the coefficients $\omega_{N, \mathbf{R}, \chi}$ are determined by the precise microscopic details. However, while the mutual braiding processes of Ising anyons bound to electrons can change the internal degrees of freedom $\chi$, the transfer of electrons in or out of the disk is forbidden by the large charging energy $E_{\mathrm{disk}} \gg \tilde{t}$, which means that the electron number $N$ and the total anyon content $a$ of the disk are fixed. Since there are three kinds of anyons in the non-Abelian Kitaev spin liquid, the total anyon content can take three different values: trivial boson ($a = 1$), Majorana fermion ($a = \psi$), or Ising anyon ($a = \sigma$). Also, the possible values of $a$ for each $N$ are restricted by the fusion rules of the non-Abelian Kitaev spin liquid~\cite{Kitaev-2006}. In particular, the disk is bosonic ($a = 1$) for $N = 0$, it hosts a net Ising anyon ($a = \sigma$) for odd $N$, while it may be bosonic ($a = 1$) or fermionic ($a = \psi$) for even $N > 0$. Hence, by tuning the electron number $N = N_{\mathrm{disk}}$ of the disk via the gate voltage $V_{\mathrm{disk}}$, different types of anyons can be stabilized in the disk, including a trivial boson and a non-Abelian Ising anyon in the simplest cases of $N = 0$ and $N = 1$, respectively.

\emph{Anyon detection.}---The stabilization of anyons in the disk region can be readily confirmed by measuring the electrical conductance between the two metallic leads connected to the semiconductor [see Fig.~\ref{fig-1}(a)]. Since the electron number inside the disk region is fixed, while the electron number of the ring region can fluctuate between $0$ and $1$, the path of each electron traveling between the two leads is entirely inside the ring region. In the limit of $W \gg t$ and $\tilde{t} \ll K$, this electron in the ring also has an Ising anyon bound to it, which means that the interference between electron paths above and below the disk is sensitive to anyons inside the disk.

To obtain the conductance between the two leads, we must first consider the relevant low-energy states of $H_{\mathrm{structure}}$ for $N_{\mathrm{disk}} = N$ and $N_{\mathrm{ring}} = 1$. For $W \to \infty$, the electrons are localized at specific dopant sites in both the disk and the ring, each of them binding an Ising anyon, and a general low-energy eigenstate can be written as $| \mathbf{r}; N, \mathbf{R}, \chi \rangle$, where $\mathbf{r}$ is the position of the electron in the ring, while $\mathbf{R}$ and $\chi$ have the same meaning as earlier. At finite $W$, since electron hopping inside the ring or electron tunneling between the ring and the leads does not affect the internal properties $N$, $\mathbf{R}$, and $\chi$ of the disk, the low-energy eigenstates that contribute to the conductance must take the form
\begin{equation}
| \Psi_{N, n} \rangle = \frac{1} {\sqrt{\mathcal{N}}} \sum_{\mathbf{r} \in \mathbb{D}_{\mathrm{ring}}} \sum_{\mathbf{R}, \chi} \theta_{\mathbf{r}, n} \, \omega_{N, \mathbf{R}, \chi} \, | \mathbf{r}; N, \mathbf{R}, \chi \rangle \label{eq-Psi}
\end{equation}
in terms of a generic label $n$, where $\mathcal{N}$ is the number of dopant sites in the ring, $\mathbf{r} \in \mathbb{D}_{\mathrm{ring}}$. If these dopant sites form a closed loop of length $\ell$ [see Fig.~\ref{fig-1}(a)] with an approximately uniform spacing $d = \ell / \mathcal{N}$ and electron hopping amplitude $\tilde{t}$ between them, the coefficient $\theta_n (x) \equiv \theta_{\mathbf{r}, n}$ is a slowly-varying $O(1)$ function of the electron position $0 \leq x \leq \ell$ along the loop and is governed by the effective low-energy continuum Hamiltonian $\mathcal{H} = -\tilde{t} d^2 \nabla^2$. Importantly, the corresponding boundary condition, $\theta_n (0) = e^{i \varphi} \, \theta_n (\ell)$, reflects both the magnetic flux inside the loop via the Aharonov-Bohm effect picked up by the electron itself, and the total anyon content $a$ of the disk via the braiding rules of the non-Abelian Kitaev spin liquid, as seen by the Ising anyon attached to the electron. In particular, $\varphi = -\phi + \vartheta_a$, where $\phi = e \Phi / \hbar = \sum_{\langle \mathbf{r}, \mathbf{r}' \rangle} A_{\mathbf{r}, \mathbf{r}'}$ [see Eq.~(\ref{eq-H-semi})] is the dimensionless magnetic flux, while the braiding phases $\vartheta_a$ for $a = 1, \psi, \sigma$ are given by~\cite{Kitaev-2006} $\vartheta_1 = 0$, $\vartheta_{\psi} = \pi$, and $\vartheta_{\sigma} = \pm \pi/4$, with the sign of $\vartheta_{\sigma}$ determined by the magnetic-field direction~\cite{Footnote-1}. The allowed wave functions corresponding to $| \Psi_{N, n} \rangle$ are then given by $\theta_n (x) = \exp[i (2\pi n - \varphi) x / \ell]$, and the relative energies of $| \Psi_{N, n} \rangle$ with respect to $| \Omega_N \rangle$ take the form $\varepsilon_n = \delta E + \tilde{t} d^2 (2\pi n - \varphi)^2 / \ell^2$, where $n$ is an integer phase winding number, while $\delta E$ is the energy difference between the $N_{\mathrm{ring}} = 1$ and $N_{\mathrm{ring}} = 0$ ground states or, in other words, a finite shift from the Coulomb-blockade peak that can be tuned via the gate voltage $V_{\mathrm{ring}}$.

\begin{figure}[t]
\includegraphics[width=0.95\columnwidth]{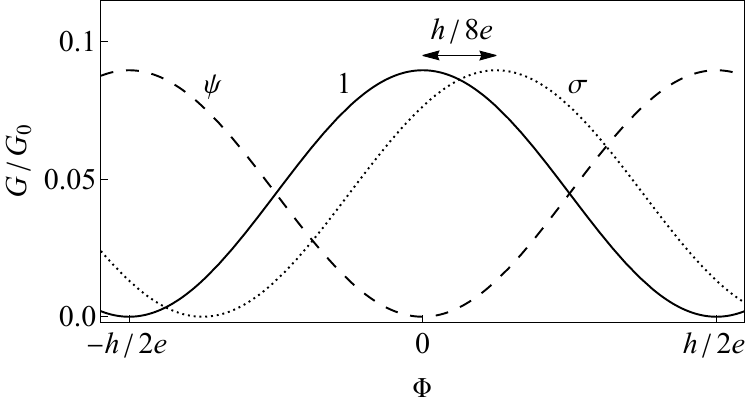}
\caption{Electrical conductance $G$ of the ring region, in units of the conductance quantum $G_0 = 2e^2 / h$, against the magnetic flux $\Phi$ if the disk region inside contains a trivial boson $1$ (solid line), a Majorana fermion $\psi$ (dashed line), or an Ising anyon $\sigma$ (dotted line). The conductance is computed via Eq.~(\ref{eq-G}) for $\rho T_0^2 = 10$ meV, $\tilde{t} = 1$ meV, $\delta E = 10$ $\mu$eV, $\Gamma_{\mathrm{L}} = \Gamma_{\mathrm{R}} = 0.1$, and $\mathcal{N} = \ell / d = 100$.} \label{fig-2}
\end{figure}

Next, we assume without loss of generality that an electron from each lead $\Lambda = \mathrm{L}, \mathrm{R}$ can directly tunnel to a single dopant site $\mathbf{r}_{\Lambda} \in \mathbb{D}_{\mathrm{ring}}$ along the edge of the spin liquid [see Fig.~\ref{fig-1}(a)] with tunneling amplitude $T_{\Lambda, \mathbf{r}_{\Lambda}} = T_0$ [see Eq.~(\ref{eq-H-tunnel})]. If $\delta E > 0$ such that $| \Omega_N \rangle$ is the overall ground state of $H_{\mathrm{structure}}$, the conductance between the two leads at zero temperature from Fermi's golden rule is then $G = 2\pi (e^2 / \hbar) \rho^2 \sum_{\mu, \nu} |S_{\mu, \nu}|^2$, where $\rho$ is the density of states in each lead, while
\begin{align}
S_{\mu, \nu} &= T_0^2 \sum_n \varepsilon_n^{-1} \langle \Omega_N | c_{\mathbf{r}_{\mathrm{R}}, \mu}^{\phantom{\dag}} | \Psi_{N, n} \rangle \langle \Psi_{N, n} | c_{\mathbf{r}_{\mathrm{L}}, \nu}^{\dag} | \Omega_N \rangle \nonumber \\
&= T_0^2 M_{\mathrm{R}, \mu}^{\phantom{*}} M_{\mathrm{L}, \nu}^{*} \sum_n \frac{\theta_{\mathbf{r}_{\mathrm{R}}, n}^{\phantom{*}} \theta_{\mathbf{r}_{\mathrm{L}}, n}^{*}} {\mathcal{N} \varepsilon_n} \label{eq-S}
\end{align}
is a spin-dependent cotunneling amplitude in terms of the matrix elements $M_{\Lambda, \mu} = \langle N, \mathbf{R}, \chi | c_{\mathbf{r}_{\Lambda}, \mu} | \mathbf{r}_{\Lambda}; N, \mathbf{R}, \chi \rangle \sim 1$ that do not depend on the internal properties $N$, $\mathbf{R}$, and $\chi$ of the disk~\cite{Footnote-2}. Finally, if the dopant sites $\mathbf{r}_{\mathrm{L}}$ and $\mathbf{r}_{\mathrm{R}}$ correspond to positions $x = 0$ and $x = \ell / 2$ along the loop of dopant sites around the ring [see Fig.~\ref{fig-1}(a)], the conductance takes the form
\begin{align}
G &= \frac{2\pi e^2} {\hbar} \, \frac{\rho^2 T_0^4 \Gamma_{\mathrm{L}} \Gamma_{\mathrm{R}}} {\mathcal{N}^2 \delta E^2} \left| \sum_{n = -\infty}^{+\infty} \frac{e^{i (2\pi n - \varphi) / 2}} {1 + \lambda^{-2} (2\pi n - \varphi)^2} \right|^2 \nonumber \\
&= \frac{2\pi e^2} {\hbar} \, \frac{\rho^2 T_0^4 \Gamma_{\mathrm{L}} \Gamma_{\mathrm{R}}} {\tilde{t} \, \delta E} \, \frac{\sinh^2 (\lambda / 2) \cos^2 (\varphi / 2)} {[\cosh \lambda - \cos \varphi]^2} \label{eq-G}
\end{align}
in terms of $\Gamma_{\Lambda} = \sum_{\mu} | M_{\Lambda, \mu} |^2$, where $\lambda = (\ell / d) (\delta E / \tilde{t})^{1/2}$ is roughly the number of eigenstates $| \Psi_{N, n} \rangle$ inside the lowest energy range of width $\delta E$, while $\varphi = -\phi + \vartheta_a$ is the effective Aharonov-Bohm phase including the dimensionless magnetic flux $\phi = e \Phi / \hbar$ and the anyonic braiding phase $\vartheta_a$. As shown by Fig.~\ref{fig-2}, the conductance $G$ exhibits Aharonov-Bohm oscillations with periodicity $\Delta \Phi = h / e$ in the magnetic flux $\Phi$, and the total anyon content $a$ of the disk region is reflected in a shift of these oscillations. In particular, whereas the maxima of $G$ are at integer multiples of $h / e$ for $a = 1$, they are at half-integer multiples for $a = \psi$ and shifted by $h / (8e)$ with respect to integer multiples for $a = \sigma$. We remark that the sign of this shift can be switched by reversing the in-plane magnetic field that does not contribute to the flux $\Phi$.

\emph{Discussion.}---The most promising candidate system for the non-Abelian Kitaev spin liquid in our setup is $\alpha$-RuCl$_3$~\cite{Plumb-2014, Sandilands-2015, Sears-2015, Majumder-2015, Johnson-2015, Sandilands-2016, Banerjee-2016, Banerjee-2017, Do-2017} with an appropriate in-plane magnetic field~\cite{Kubota-2015, Leahy-2017, Sears-2017, Wolter-2017, Baek-2017, Banerjee-2018, Hentrich-2018, Jansa-2018, Kasahara-2018, Widmann-2019, Balz-2019, Yamashita-2020, Czajka-2021, Yokoi-2021, Bruin-2022, Czajka-2023}, while the monolayer semiconductor coupled to it could be a transition-metal dichalcogenide like MoS$_2$ or WSe$_2$~\cite{Manzeli-2017}. In these two-dimensional materials, arrays of electron dopants can be created with atomistic precision using the focused electron beam of a scanning transmission electron microscope~\cite{Dyck-2019}.

For stabilizing anyons in the disk region, the effective electron hopping amplitude $\tilde{t}$ must be much smaller than both the Kitaev exchange $K$ and the charging energy $E_{\mathrm{disk}}$, which implies that $\tilde{t}$ is around $1$ meV or below. At the same time, these anyons can only be detected in the conductance of the ring region if $\lambda = (\ell / d) (\delta E / \tilde{t})^{1/2} \lesssim 10$, otherwise the interference effect is exponentially suppressed [see Eq.~(\ref{eq-G})]. Thus, for ring length $\ell \sim 500$ nm and dopant spacing $d \sim 5$ nm, the energy shift with respect to the Coulomb-blockade peak is limited to $\delta E \lesssim 10$ $\mu$eV, which in turn requires the gate voltage $V_{\mathrm{ring}}$ to be tuned with $10$ $\mu$V accuracy and the temperature to be in the $100$ mK range or below. Similarly, gapless excitations along the edge of the spin liquid, which were ignored in our calculation, may destroy the interference if $\delta E$ is much larger than their energy spacing, $\epsilon \sim K a_0 / L$. Thus, for Kitaev exchange $K \sim 10$ meV and lattice constant $a_0 \sim 1$ nm, the edge length is constrained to be $L \lesssim 1$ $\mu$m. Finally, to observe a shift of $h / (8e)$ in the Aharonov-Bohm oscillations of the conductance for a ring length $\ell \sim 500$ nm, an out-of-plane magnetic field of around $100$ mT is necessary. We emphasize, however, that the conductance measured at zero flux is already sensitive to the total anyon content of the disk region.

We also point out that our key ideas apply beyond the concrete setup described in this work. For example, the need for a Kondo-coupled semiconducting monolayer could be circumvented by directly doping $\alpha$-RuCl$_3$ with mobile charge carriers~\cite{Biswas-2019, Wang-2020}. Alternatively, instead of introducing an array of electron dopants into a monolayer semiconductor, one could use a twisted bilayer semiconductor to produce slow electron dynamics in the sub-meV range~\cite{Wu-2019, Devakul-2021}.

\emph{Acknowledgments.}---We thank Jason Alicea, Tom Berlijn, Matthew Brahlek, Yong Chen, Stephen Jesse, Kai Klocke, and Alan Tennant for enlightening discussions and helpful comments on the manuscript. This material is based upon work supported by the U.S. Department of Energy, Office of Science, National Quantum Information Science Research Centers, Quantum Science Center.





\begin{references}
\bibitem{Footnote-1} Note that $\vartheta_{\sigma}$ also depends on the fusion channel between the Ising anyons in the disk and the ring. However, we can assume that this fusion channel is bosonic in the lowest-energy states because the fermionic fusion channel requires another fermion at the edge of the spin liquid that costs finite energy.
\bibitem{Footnote-2} Strictly speaking, these matrix elements exhibit an orthogonality catastrophe due to the creation or annihilation of an Ising anyon at the gapless edge of the spin liquid. In practice, however, the corresponding factors $(L/a_0)^{-1/16}$ in the matrix elements are $O(1)$ if the edge length is not too large compared to the lattice constant, $L/a_0 \lesssim 10^6$.
\end{references}
\end{document}


\title{Gate-controlled anyon generation and detection in Kitaev spin liquids \\ (Supplemental Material)}

\author{G\'abor B. Hal\'asz}
\affiliation{Materials Science and Technology Division, Oak Ridge National Laboratory, Oak Ridge, TN 37831, USA}
\affiliation{Quantum Science Center, Oak Ridge, TN 37831, USA}


\maketitle

\renewcommand{\thefigure}{S\arabic{figure}}
\renewcommand{\theequation}{S\arabic{equation}}


\section{Isolated Kondo impurity in a non-Abelian Kitaev spin liquid}

Here we consider a single impurity spin $\vec{\Sigma}$ at position $\mathbf{r} = \mathbf{r}_0$ that is Kondo coupled to a non-Abelian Kitaev spin liquid in the presence of a Zeeman field $\vec{h}$. The relevant Hamiltonian is given by [see Eqs.~(3), (4), and (5) in the main text]
\begin{equation}
\tilde{H} = H_{\mathrm{Kitaev}} + J \left( \vec{\sigma}_{\mathbf{r}_0} \cdot \vec{\Sigma} \right) - \vec{h} \cdot \vec{\Sigma}, \qquad H_{\mathrm{Kitaev}} = -K \sum_{\langle \mathbf{r}, \mathbf{r}' \rangle_{\alpha}} \sigma_{\mathbf{r}}^{\alpha} \sigma_{\mathbf{r}'}^{\alpha} - \kappa K \sum_{\langle \mathbf{r}, \mathbf{r}', \mathbf{r}'' \rangle_{\alpha \beta}} \sigma_{\mathbf{r}}^{\alpha} \sigma_{\mathbf{r}'}^{\gamma} \sigma_{\mathbf{r}''}^{\beta}, \label{eq-H}
\end{equation}
where $\vec{\Sigma} = (\Sigma^x, \Sigma^y, \Sigma^z)$, $\vec{\sigma}_{\mathbf{r}} = (\sigma_{\mathbf{r}}^x, \sigma_{\mathbf{r}}^y, \sigma_{\mathbf{r}}^z)$, and the spin components $\Sigma^{\alpha}$ and $\sigma_{\mathbf{r}}^{\alpha}$ with $\alpha = x,y,z$ all square to unity. We aim to find the critical Kondo coupling $J$ beyond which an Ising anyon is bound to the impurity spin.

\begin{figure}[h]
\includegraphics[width=0.9\columnwidth]{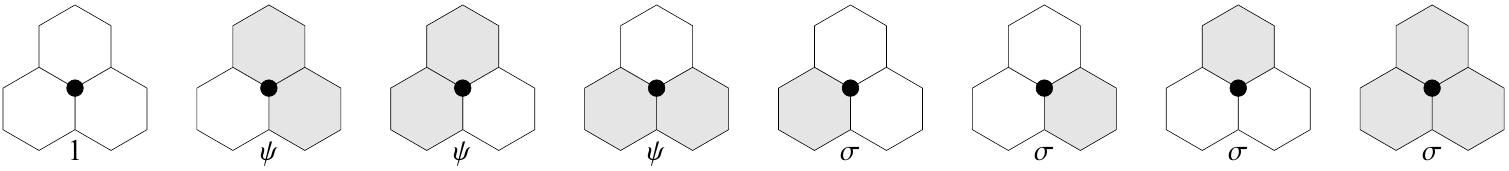}
\caption{Flux sectors of the Kitaev model coupled by the Kondo coupling around the impurity position $\mathbf{r}_0$ (black dot). For each flux sector, flux eigenvalues $+1$ and $-1$ are denoted by white and gray color, respectively, while the label $a = 1, \psi, \sigma$ specifies the total anyon content of the impurity region. Note that all fluxes not shown have eigenvalues $+1$, except for one far-away flux with eigenvalue $-1$ in the $a = \sigma$ cases.} \label{fig-S1}
\end{figure}

For sufficiently small $J \lesssim K$, we can treat the Kondo coupling term $\propto J$ perturbatively by projecting it onto the low-energy manifold of the Kitaev Hamiltonian $H_{\mathrm{Kitaev}}$. Since the fermion gap is between $K$ and $2K$ while the flux gap is between $0.25K$ and $0.4K$ for any $0.1 \leq \kappa \leq 0.2$, it is consistent to ignore the fermion excitations but keep the flux excitations. Given that the Kitaev spin $\vec{\sigma}_{\mathbf{r}_0}$ coupled to the impurity spin $\vec{\Sigma}$ only anticommutes with three flux operators, the low-energy manifold is then $16$ dimensional, and its states can be written as $| F \rangle \otimes | \Sigma \rangle$, where $| F \rangle$ are the fermion vacuum states of the flux sectors depicted in Fig.~\ref{fig-S1}, while $| \Sigma \rangle$ are the two possible states of the impurity spin (e.g., ``up'' and ``down''). Since $| F \rangle$ are eigenstates of $H_{\mathrm{Kitaev}}$, the matrix elements of $\tilde{H}$ within this low-energy manifold can then be calculated via the exact solution of $H_{\mathrm{Kitaev}}$. In practice, we compute these matrix elements numerically for a $20 \times 20$ system with periodic boundary conditions, and also verify that they remain essentially the same if the system size is reduced to $16 \times 16$ or $12 \times 12$.

\begin{figure}[t]
\includegraphics[width=0.9\columnwidth]{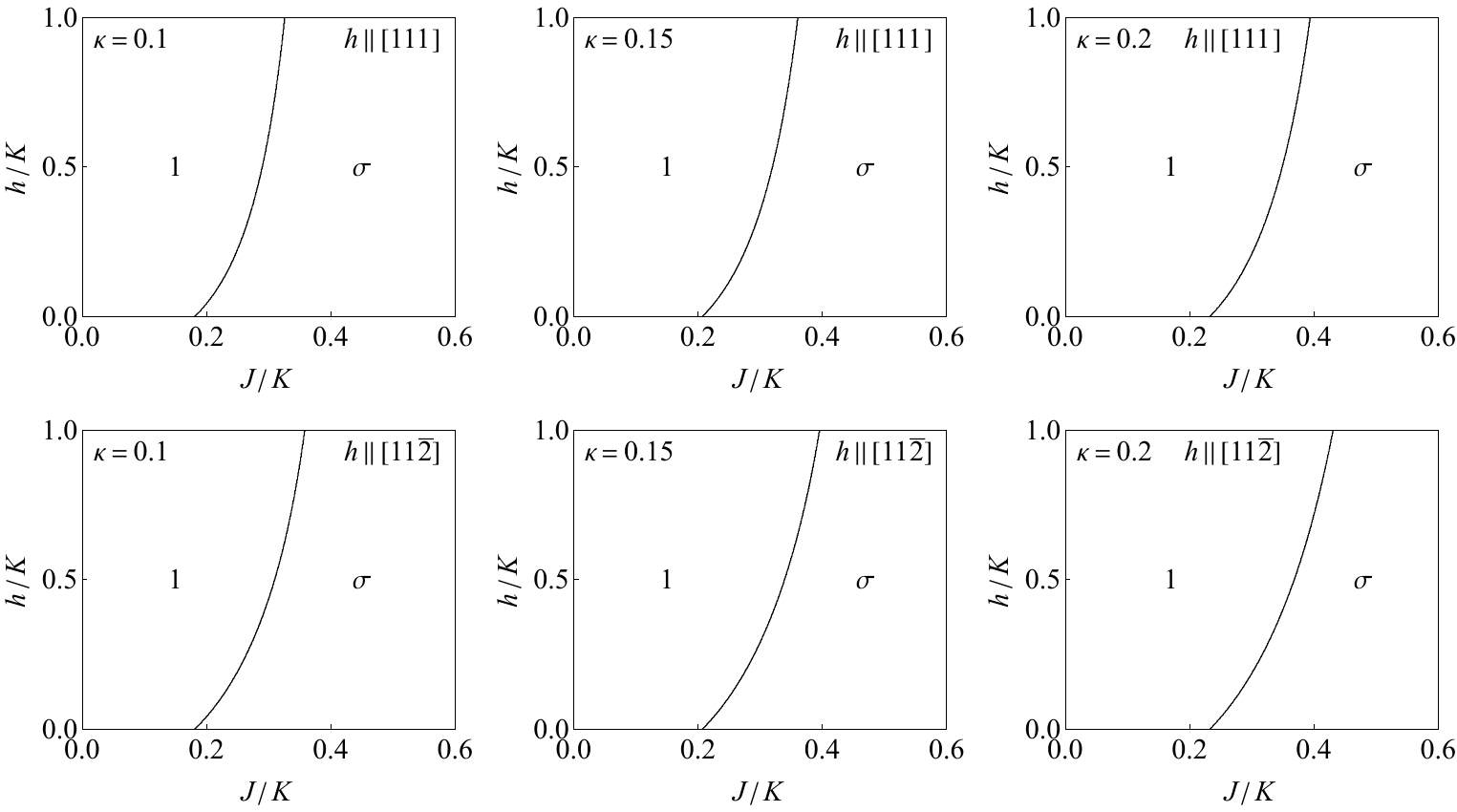}
\caption{Total anyon content $a = 1, \psi, \sigma$ of the impurity region in the ground state of the Hamiltonian $\tilde{H}$ [see Eq.~(\ref{eq-H})] against the Kondo coupling $J$ and the Zeeman field $h \equiv |\vec{h}|$ for out-of-plane ($\vec{h} \parallel [111]$) and in-plane ($\vec{h} \parallel [11\bar{2}]$) field directions (top and bottom, respectively) at $\kappa = 0.1$ (left), $\kappa = 0.15$ (center), and $\kappa = 0.2$ (right).} \label{fig-S2}
\end{figure}

Importantly, each Kitaev eigenstate $| F \rangle$ has a well-defined total anyon content $a$ around the impurity spin (see Fig.~\ref{fig-S1}), and the low-energy manifold thus fragments into three disconnected submanifolds corresponding to $a = 1, \psi, \sigma$. The states $| F \rangle$ with $a = 1$ and $a = \psi$ have even and odd fermion parities, respectively, while the states with $a = \sigma$ have no well-defined fermion parity due to a zero-energy fermion mode shared by the flux bound to the impurity spin and another flux far away. For each set of parameters in Eq.~(\ref{eq-H}), we can then diagonalize the projection of the Hamiltonian $\tilde{H}$ into each submanifold, and determine which submanifold contains the ground state~\cite{Footnote}. The results in Fig.~\ref{fig-S2} show that the ground state switches from $a = 1$ to $a = \sigma$ at a critical Kondo coupling in the range of $0.2K \lesssim J \lesssim 0.4K$ for any $0.1 \leq \kappa \leq 0.2$ and for both out-of-plane ($\vec{h} \parallel [111]$) and in-plane ($\vec{h} \parallel [11\bar{2}]$) directions of the Zeeman field as long as the magnitude of the field is sufficiently small ($|\vec{h}| \lesssim K$). In other words, we find that an Ising anyon is bound to the impurity spin beyond this critical Kondo coupling.

